\documentclass[12pt]{article}
\usepackage{amsmath,amssymb,latexsym,epsf,epsfig,graphicx,tikz}

\begin{document}

\newcommand{\sect}[1]{\setcounter{equation}{0}\section{#1}}
\renewcommand{\theequation}{\thesection.\arabic{equation}}
\newcommand{\prt}{\partial}
\newcommand{\II}{\mbox{${\mathbb I}$}}
\newcommand{\CC}{\mbox{${\mathbb C}$}}
\newcommand{\RR}{\mbox{${\mathbb R}$}}
\newcommand{\QQ}{\mbox{${\mathbb Q}$}}
\newcommand{\ZZ}{\mbox{${\mathbb Z}$}}
\newcommand{\NN}{\mbox{${\mathbb N}$}}
\def\G{\mathbb G}
\def\UU{\mathbb U}
\def\S{\mathbb S}
\def\T{\mathbb T}
\def\HH{\mathbb H}
\def\tS{\widetilde{\mathbb S}}
\newcommand{\mB}{{\mathbb B}}
\newcommand{\mA}{{\mathbb A}}
\newcommand{\mC}{{\mathbb C}}

\def\V{\mathbb V}
\def\tV{\widetilde{\mathbb V}}
\newcommand{\D}{{\mathbb D}}
\def\hint{H_{\rm int}}
\def\R{{\cal R}}

\newcommand{\rd}{{\rm d}}
\newcommand{\diag}{{\rm diag}}
\newcommand{\U}{{\cal U}}
\newcommand{\K}{{\mathcal K}}
\newcommand{\cP}{{\cal P}}
\newcommand{\dQ}{{\dot Q}}
\newcommand{\dS}{{\dot S}}
\newcommand{\W}{{\mathcal W}}

\newcommand{\pnf}{P^N_{\rm f}}
\newcommand{\pnb}{P^N_{\rm b}}
\newcommand{\hnf}{P^Q_{\rm f}}
\newcommand{\hnb}{P^Q_{\rm b}}

\newcommand{\ph}{\varphi}
\newcommand{\phd}{\widetilde{\varphi}} 
\newcommand{\phs}{\varphi^{(s)}}
\newcommand{\phb}{\varphi^{(b)}}
\newcommand{\phds}{\widetilde{\varphi}^{(s)}}
\newcommand{\phdb}{\widetilde{\varphi}^{(b)}}
\newcommand{\lambdad}{\widetilde{\lambda}}
\newcommand{\tx}{\widetilde{x}} 
\newcommand{\td}{\widetilde{d}} 
\newcommand{\etat}{\widetilde{\eta}}
\newcommand{\phl}{\varphi_{i,L}}
\newcommand{\phr}{\varphi_{i,R}}
\newcommand{\phz}{\varphi_{i,Z}}
\newcommand{\mum}{\mu_{{}_-}}
\newcommand{\mup}{\mu_{{}_+}}
\newcommand{\mupm}{\mu_{{}_\pm}}
\newcommand{\muv}{\mu_{{}_V}}
\newcommand{\mua}{\mu_{{}_A}}
\newcommand{\wt}{\hat{t}}

\def\a{\alpha}
 
\def\A{\mathcal A} 
\def\H{\mathcal H} 
\def\U{\mathcal U} 
\def\E{\mathcal E} 
\def\C{\mathcal C} 
\def\L{\mathcal L} 
\def\M{\mathcal M} 
\def\O{\mathcal O}
\def\I{\mathcal I}
\def\Z{\mathcal Z} 
\def\der{\partial }
\def\mis{{\frac{\rd k}{2\pi} }}
\def\ri{{\rm i}}
\def\xt{{\widetilde x}}
\def\ft{{\widetilde f}}
\def\gt{{\widetilde g}}
\def\qt{{\widetilde q}}
\def\tt{{\widetilde t}}
\def\tmu{{\widetilde \mu}}
\def\prt{{\partial}}
\def\tr{{\rm Tr}}
\def\inc{{\rm in}}
\def\out{{\rm out}}
\def\Li{{\rm Li}}
\def\e{{\rm e}}
\def\eps{\varepsilon}
\def\k{\kappa}
\def\v{{\bf v}}
\def\ebf{{\bf e}}
\def\abf{{\bf A}}
\def\fa{{\mathfrak a}} 

%%%%%%%%%%%%%%%%%%% INIZIO %%%%%%%%%%%%%%%%%%%%%%%

%%%%%%%% 
\newcommand{\finprf}{\null \hfill {\rule{5pt}{5pt}}\\[2.1ex]\indent}

%%%%%%%%%%%%%%%%%%%%%%%
\pagestyle{empty}
\rightline{January 2016}
%\rightline{Preliminary Draft}

\bigskip 

\begin{center}
{\Large\bf Non-equilibrium Current Cumulants\\  
and Moments with a Point-like Defect}
\\[2.1em]

\bigskip

{\large Mihail Mintchev}\\ 
\medskip 
{\it  
Istituto Nazionale di Fisica Nucleare and Dipartimento di Fisica, Universit\`a di
Pisa, Largo Pontecorvo 3, 56127 Pisa, Italy}
\bigskip 

{\large Luca Santoni}\\ 
\medskip 
{\it  
Scuola Normale Superiore and Istituto Nazionale di Fisica Nucleare, Piazza dei Cavalieri 7, 56126 Pisa, Italy}
\bigskip 

{\large Paul Sorba}\\ 
\medskip 
{\it  
LAPTh, Laboratoire d'Annecy-le-Vieux de Physique Th\'eorique, 
CNRS, Universit\'e de Savoie,   
BP 110, 74941 Annecy-le-Vieux Cedex, France}
\bigskip 
\bigskip 
\bigskip 

\end{center}
\begin{abstract} 
\bigskip 

We derive the exact $n$-point current expectation values in the Landauer-B\"uttiker 
non-equilibrium steady state of a multi terminal system with star graph geometry 
and a point-like defect localised in the vertex. The current cumulants are extracted from 
the connected correlation functions and the cumulant generating function is established. 
We determine the moments, show that the associated moment problem has a unique 
solution and reconstruct explicitly the corresponding probability distribution. The basic building 
blocks of this distribution are the probabilities of particle emission and absorption from 
the heat reservoirs, driving the system away from equilibrium. We 
derive and analyse in detail these probabilities, showing that they fully describe the 
quantum transport problem in the system.

\end{abstract}
\bigskip 
\medskip 
\bigskip 

\vfill
\rightline{IFUP-TH 1/2016}
\rightline{LAPTH-069/15}
\newpage
\pagestyle{plain}
\setcounter{page}{1}

%%%%%%%%%%%%%%%%%%%%%%%%%%%%%%%%
\section{Introduction} 
\medskip

Current fluctuations represent a fundamental characteristic feature of non-equilibrium 
quantum transport. A complete information about these fluctuations  
is provided by the cumulants $\C_n$ of the particle 
current, which generalise the quadratic noise fluctuations 
to $n\geq 3$ currents. For this reason the study of the sequence $\{\C_n\, :\, n=1,2,...\}$ 
attracted much attention in last two decades. Following the fundamental work of Khlus \cite{K-87} and 
Levitov, Lesovik and Chtchelkatchev \cite{L-89}-\cite{LC-03}, there has been a series of 
contributions studying various systems \cite{LLL-96}-\cite{Bee-15} and different non-equilibrium situations. 
Several examples are discussed in the proceedings \cite{N-03} as well as in the review 
papers \cite{BB-00}-\cite{LS-14} and the references therein. 

\begin{figure}[h]
\begin{center}
\begin{picture}(600,120)(35,275) 
\includegraphics[scale=0.8]{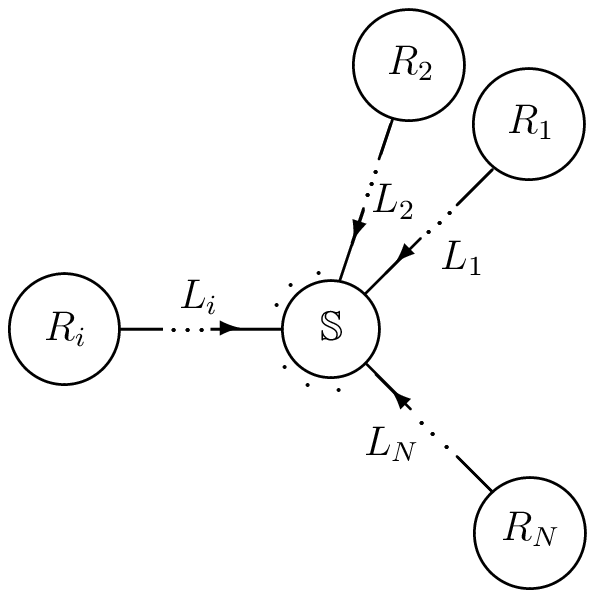}
\end{picture} 
\end{center}
\caption{$N$-terminal junction with scattering matrix $\S$.} 
\label{fig1}
\end{figure} 

In this paper we consider the particle current fluctuations for general class of point-like defects. More precisely 
we investigate a system with $N$ terminals with the geometry of a star graph 
as shown in Fig. \ref{fig1}. Each of the $N$ semi-infinite leads 
is attached at infinity to the a heat reservoir $R_i$ with (inverse) temperature $\beta_i$ and 
chemical potential $\mu_i$. The defect, which drives the system out of equilibrium, 
is localised in the vertex of the graph and is described by a $N\times N$ 
unitary scattering matrix $\S$. Although relatively simple, the system in Fig. \ref{fig1} 
represents a remarkable laboratory for studying 
a large class of intriguing quantum phenomena. The deep relation between the particle  
density cumulants and the R\'enyi entanglement entropies in an equilibrium configuration 
at zero temperature has been investigated in \cite{CMV-12}. The study of the first cumulant of the 
particle and heat currents away from equilibrium  
shows \cite{MSS-14} that the junction transforms heat to chemical potential energy and vice versa, 
depending on the parameters of the heat baths. 
The explicit form of the second cumulant $\C_2$ in the scale invariant limit 
reveals \cite{MSS-15} the existence of nonlinear 
effects, which lead to reduction or enhancement of the particle and heat noises 
in certain ranges of the chemical potentials. In what follows we pursue further 
the investigation of the junction in Fig. \ref{fig1}, adopting the following strategy. 
In a quantum field theory framework we first derive in explicit and closed form 
the particle current cumulants $\C_n$ for generic $n$. We use this information for reconstructing 
both the cumulant and the moment generating functions. From these data we finally recover 
the associated probability distribution, which captures the microscopic characteristic features of the system. 
In this respect we explicitly determine the probability $p_{ij}$ of a particle with given energy to be emitted from the heat 
reservoir $R_i$ and absorbed by $R_j$. It turns out that $p_{ij}$ are nontrivial in both 
cases $i=j$ and $i\not=j$. Our investigation covers all point-like defects in the vertex, 
which are compatible with a unitary time evolution in the bulk generated by the Schr\"odinger Hamiltonian. We call 
these systems Schr\"odinger junctions. 

The paper is organised as follows. In the next section we first recall the form of the particle current 
in the presence of a point-like defect. Afterwards we derive the exact connected $n$-point 
current correlation functions in the Landauer-B\"uttiker 
non-equilibrium steady state. The zero-frequency limit of these functions defines the cumulants $\C_n$. 
The cumulant generating function $\C(\lambda )$ in the general case with $N$-terminals 
is reconstructed in section 3. We establish here also the explicit form of $\C(\lambda )$ 
in the scale invariant limit and describe the main properties of the 
cumulants $\C_3$ and $\C_4$ in this regime. The last part of section 3 
provides a comparison with other results in the subject. 
In section 4 we derive the moments in a single energy channel, 
the associated probability distribution and the probabilities $p_{ij}$ 
mentioned above. Our conclusions are collected in section 5, whereas the appendices 
contain some technical details.  

\bigskip

\section{Current correlation functions and cumulants} 

\subsection{Particle current in the presence of defect}

\medskip 

The system in Fig. \ref{fig1} is localised on a star graph 
with coordinates $\{(x,i)\, ,:\, x\leq 0,\, i=1,...,N\}$, where $x$ denotes 
the distance from the vertex and $i$ labels the leads. The main 
object of our investigation is the particle current $j(t,x,i)$ flowing along the 
leads of the junction. In order to determine $j(t,x,i)$ one should fix both the 
dynamics in the bulk and the boundary conditions in the vertex of the graph. 
We consider\footnote{The natural units $\hbar =c=k_{\rm B}=1$ are adopted 
throughout the paper.}
\begin{equation}
\left (\ri \prt_t +\frac{1}{2m} \prt_x^2\right )\psi (t,x,i) = 0\, ,  
\label{eqm1}
\end{equation} 
with the boundary condition 
\begin{equation} 
\lim_{x\to 0^-}\sum_{j=1}^N \left [\eta (\II-\UU)_{ij} +\ri (\II+\UU)_{ij}\prt_x \right ] \psi (t,x,j) = 0\, , 
\label{bc1} 
\end{equation} 
where $\UU$ is a $N\times N$ unitary matrix and $\eta \in \RR$ is a 
parameter with dimension of mass. Equation (\ref{bc1}) parametrises all 
self-adjoint extensions of the bulk Hamiltonian $-\prt_x^2$ to the whole 
graph and gives rise to non-trivial one-body 
interactions, which are described by the scattering matrix \cite{ks-00}-\cite{k-08} 
\begin{equation} 
\S(k) = 
-\frac{[\eta (\II - \UU) - k(\II+\UU )]}{[\eta (\II - \UU) + k(\II+\UU)]} \, ,   
\label{S1}
\end{equation} 
$k$ being the particle momentum. We stress that the scattering matrices (\ref{S1}) 
provide a physical description of all point-like contact interactions among the leads, 
which are compatible with a unitary time evolution in the bulk of the system. This is  
the fundamental requirement selecting the class of defects considered in this paper. 

The solution of (\ref{eqm1},\ref{bc1}) is given by 
\begin{equation} 
\psi (t,x,i)  = \sum_{j=1}^N \int_{0}^{\infty} \frac{dk}{2\pi } \e^{-\ri \omega (k)t}\, 
\chi_{ij}(k;x) a_j (k) \, , \qquad  \omega(k) = \frac {k^2}{2m} \, .
\label{psi1} 
\end{equation} 
where 
\begin{equation}
\chi(k;x) = \e^{-\ri k x}\, \II + \e^{\ri k x}\, \S(-k)  
\label{chi}
\end{equation} 
and $\{a_i(k),\, a^*_i(k)\, :\, k \geq 0,\, i=1,...,N\}$ 
generate the standard canonical anticommutation relation algebra.  
With the above definitions the particle current takes the form\footnote{The ${}^*$ stands for Hermitian conjugation.}
\begin{eqnarray}
j(t,x,i)= \frac{\ri}{2m} \int_0^\infty \frac{\rd k}{2\pi} \int_0^\infty \frac{\rd p}{2\pi}\,  
\e^{\ri t [\omega(k) - \omega(p)]} \qquad \qquad \qquad 
\nonumber \\ 
\times \sum_{l,m=1}^N a^*_l(k) \Bigl \{
\chi^*_{li}(k;x) \left [\der_{x} \chi_{i m}\right ](p;x) - 
\left [\der_{x} \chi^*_{li}\right ](k;x) \chi_{im}(p;x) \Bigr \}a_m(p) \, .
\label{curr1}
\end{eqnarray} 
Using the orthogonality and completeness of the system $\{\chi(k;x)\,:\, k\geq 0\, ,x\geq 0\}$ one 
can prove \cite{M-11} the operator Kirchhoff rule 
\begin{equation} 
\sum_{i=1}^N j(t,0,i) = 0\, , 
\label{KR}
\end{equation}
which is a simple but fundamental feature of the currents flowing in the junction. 

Besides the particle current operator, we have to fix also the state for evaluating the 
current expectation values. The physical setting, presented in 
Fig. \ref{fig1}, is nicely described by the Landauer-B\"uttiker (LB) \cite{L-57, B-86} 
non-equilibrium steady state $\Omega_{\beta, \mu}$, defined in terms of 
$(\beta_i,\mu_i)$ and $\S(k)$. A simple and intuitive way \cite{M-11} to 
construct this state is to use the scattering matrix $\S(k)$ in order to 
extend the tensor product of Gibbs states, relative to the 
reservoirs $R_i$ at the level of asymptotic incoming fields, to the outgoing fields.  
The state, obtained in this way, has both realistic physical properties \cite{L-57, B-86} 
and interesting mathematical structure \cite{M-11,gn,BGP-13}. 
The basic expectation values of $\{a_i(k),\, a^*_i(k)\}$ 
in $\Omega_{\beta, \mu}$, which are needed in what follows, are reported in appendix A. 
\bigskip 

\subsection{Current cumulants in the LB state} 
\medskip 

Let $L_i$ be an arbitrary but fixed lead and let us consider the $n$-point correlation function 
\begin{equation}
\W^i_n(t_1, x_1,...,t_n,x_n) = \langle j(t_1,x_1,i) \cdots j(t_n,x_n,i)  \rangle_{\beta,\mu}\, , 
\label{w1}
\end{equation} 
of the current (\ref{curr1}), where $\langle \cdots \rangle_{\beta,\mu}$ denotes the expectation 
value in the LB state $\Omega_{\beta, \mu}$. The $n$-th cumulant in $L_i$ is defined 
by the {\it connected} part of (\ref{w1}), 
\begin{equation}
\C^i_n(t_1, x_1,...,t_n,x_n) = \langle j(t_1,x_1,i) \cdots j(t_n,x_n,i)  \rangle_{\beta,\mu}^{\rm conn}\, . 
\label{c1}
\end{equation} 
{}For $n=1$ the correlators (\ref{w1},\ref{c1}) coincide and have the following well known \cite{L-57, B-86} 
time and space independent form 
\begin{equation}
\W^i_1 = \C^i_1 = \int_0^\infty \frac{\rd \omega}{2\pi} \sum_{l=1}^N 
\left (\delta_{il} -|\S_{il}(\sqrt{2m\omega})|^2\right ) d_l(\omega) \, , 
\qquad d_l(\omega)= \frac{1}{1+\e^{\beta_l (\omega -\mu_l)}}\, .   
\label{c11}
\end{equation} 

The situation complicates for $n\geq 2$. First of all the 
correlators (\ref{w1},\ref{c1}) depend on the time differences 
$\{\wt_k \equiv t_k - t_{k+1}\, :\, k=1,...,n-1\}$, which reflects 
the invariance under time translations of the LB state. Moreover, 
since the defect violates translation invariance in space, 
(\ref{w1},\ref{c1}) depend separately on all the coordinates $\{x_l\, :\, l=1,...,n\}$.  It is clear that 
dealing with this large number of variables becomes complicated with growing of $n$. 
Also, it turns out that most of them are marginal for the particle transfer we are interested in. 
One possibility \cite{N-03}-\cite{LS-14} to get rid of some space-time variables is the replacement 
\begin{equation}
j(t_l,x_l,i) \longmapsto \int_0^T \rd t_l \, j(t_l,x_l,i) \, , \qquad \forall \; l=1,...n\, , 
\label{c2}
\end{equation}
in (\ref{w1},\ref{c1}). The operation (\ref{c2}) obviously simplifies the time dependence. Instead of the $(n-1)$ time 
variables $\wt_k$, one has now only one, namely $T$. The final step in this scheme 
is to study the system for $T$ large enough. 
Unfortunately, in the presence of defect the above procedure solves the problem only partially, 
because the $x_l$-dependence persists. 

In this paper we adopt an alternative strategy, which generalises to $n\geq 3$ the definition (see e.g. \cite{BB-00}) 
of zero-frequency noise. For $n\geq 2$ we consider the Fourier transforms 
\begin{equation} 
\Z^i_n(x_1,...,x_n;\nu ) = \int_{-\infty}^{\infty} \rd \wt_1 \cdots   \int_{-\infty}^{\infty} \rd \wt_{n-1} 
\e^{\ri \nu (\wt_1+\cdots \wt_{n-1})} \Z^i_n(t_1, x_1,...,t_n,x_n)\, , \qquad \Z^i_n = \W^i_n,\, \C^i_n\, , 
\label{c3}
\end{equation}
and perform the zero-frequency limit 
\begin{equation}
\Z^i_n = \lim_{\nu \to 0^+} \Z^i_n(x_1,...,x_n;\nu ) \, .
\label{c4}
\end{equation}
We will show below that in this limit the $x_l$-dependence drops out and $\Z^i_n$ depends exclusively 
on the scattering matrix $\S$ and the heat bath parameters $(\beta_l,\mu_l)$. In fact, 
using the explicit form of the current (\ref{curr1}) and the correlation function (\ref{A6}) in appendix A, after some algebra one finds 
\begin{eqnarray}
\W^i_n=\int_0^\infty \frac{\rd \omega}{2\pi} \sum_{l_1,...,l_n=1}^N 
\left | \begin{array}{cccccccc}
\T^i_{l_1l_1}(\omega)d_{l_1}(\omega )&\T^i_{l_2l_1}(\omega)d_{l_2}(\omega )&\cdots &&  \T^i_{l_nl_1}(\omega)d_{l_n}(\omega )\\
-\T^i_{l_1l_2}(\omega)\td_{l_1}(\omega )&\T^i_{l_2l_2}(\omega)d_{l_2}(\omega )& \cdots & &\T^i_{l_nl_2}(\omega)d_{l_n}(\omega )\\
\vdots &\vdots & \vdots&& \vdots \\
-\T^i_{l_1l_n}(\omega)\td_{l_1}(\omega )&-\T^i_{l_2l_n}(\omega)\td_{l_2}(\omega ) & \cdots & &\T^i_{l_nl_n}(\omega)d_{l_n}(\omega )\\
\end{array}\right | \, ,
\label{w2}
\end{eqnarray} 
where\footnote{Here and in what follows the bar indicates complex conjugation} 
\begin{equation}
\T^i_{lm}(\omega)=\delta_{li} \delta_{mi} - \S_{li}(\sqrt{2m\omega})\, \overline{\S}_{mi}(\sqrt{2m\omega})\, , 
\label{T}
\end{equation} 
$d_l(\omega)$ is the Fermi distribution (\ref{c11}) of the reservoir $R_l$ and 
\begin{equation}
\td_l(\omega ) = 1-d_l(\omega)=\frac{\e^{\beta_l (\omega -\mu_l)}}{1+\e^{\beta_l (\omega -\mu_l)}}\, .  
\label{T1}
\end{equation}  

As expected, the connected part of (\ref{w2}) simplifies and can be 
conveniently written in terms of traces involving 
the matrices 
\begin{equation}
\mA^i \equiv \T^i \, \D\, ,\qquad \mB^i \equiv \T^i\, (\II-\D)\, ,\qquad   
\D \equiv {\rm diag}[d_1(\omega),d_2(\omega),...,d_n(\omega)] \, .
\label{m1}
\end{equation}
One finds 
\begin{eqnarray} 
\C_1^i &=& \int_0^\infty \frac{\rd \omega}{2\pi} {\rm Tr} \left [ \mA^i \right ]\, , 
\label{c51} \\
\C_n^i &=& \int_0^\infty \frac{\rd \omega}{2\pi} \sum_{\sigma \in {\cal P}_{n-1}} \, 
{\rm Tr} \left [ \mA^i \mC^i_{\sigma_1\sigma_2}\cdots \mC^i_{\sigma_{n-2}\sigma_{n-1}}\mB^i\right ]\, , 
\qquad n \geq 2\, , 
\label{c52}
\end{eqnarray}
where the sum runs over all permutations ${\cal P}_{n-1}$ of $n-1$ elements and 
\begin{equation} 
\mC^i_{\sigma_i\sigma_{i+1}} = 
\begin{cases} 
\, -\mA^i \, , & \qquad  \sigma_i < \sigma_{i+1}\, , \\
\quad \mB^i \, , & \qquad  \sigma_i > \sigma_{i+1}\, . \\ 
\end{cases} 
\label{m2}
\end{equation} 

The trace representation (\ref{c51},\ref{c52}) of the 
current cumulants with a point-like defect represents a first basic result of our study. 
It is worth stressing that the above derivation of $\C^i_n$ is purely field theoretical 
and makes no use of any kind of cumulant generating function. We will show in section 3 
that this function can be uniquely reconstructed from (\ref{c51},\ref{c52}).

\bigskip 

\subsection{The two-lead junction cumulants} 
\medskip 

In order to better illustrate the compact expressions (\ref{c51},\ref{c52}), 
it is instructive report the explicit form of the first few cumulants in the case $N=2$.  
Without loss of generality we can concentrate on the the cumulants $\C_n^1$ in the 
lead $L_1$. For notational simplicity we omit here and in what follows the 
apex $1$ in $\C_n^1$. By means of (\ref{c51})-(\ref{m2}) one gets 
\begin{eqnarray} 
\C_1 &=& \int_0^\infty \frac{\rd \omega}{2\pi}\, \tau c_1\, , 
\label{ec1}\\
\C_2 &=& \int_0^\infty \frac{\rd \omega}{2\pi}\, \tau (c_2 -\tau c_1^2)\, ,
\label{ec2}\\
\C_3 &=& \int_0^\infty \frac{\rd \omega}{2\pi}\, \tau^2 c_1(1-3c_2+2\tau c_1^2) \, ,
\label{ec3}\\
\C_4 &=& \int_0^\infty \frac{\rd \omega}{2\pi}\, \tau^2[c_2 -3c_2^2+12\tau c_1^2c_2 - 
2\tau c_1^2(2+3\tau c_1^2)]\, , 
\label{ec4}\\
\C_5 &=& \int_0^\infty \frac{\rd \omega}{2\pi}\, \tau^3 c_1[1+30c_2^2 -15c_2(1+4\tau c_1^2) + 
4\tau c_1^2(5+6\tau c_1^2)]\, , 
\label{ec5}\\
\C_6 &=& \int_0^\infty \frac{\rd \omega}{2\pi}\, \tau^3 \{c_2[1+15c_2(2c_2-1)] - 2\tau c_1^2[8+15c_2(9c_2-5)] +
\nonumber \\
&+&120 \tau^2 c_1^4(3c_2-1) -120\tau^3 c_1^6\}\, ,  
\label{ec6}
\end{eqnarray}
where the following combinations 
\begin{equation}
c_1(\omega)\equiv d_1(\omega)-d_2(\omega)\, \qquad 
c_2(\omega)\equiv d_1(\omega)+d_2(\omega)-2d_1(\omega)d_2(\omega)\, , 
\label{dc}
\end{equation}
have been introduced for convenience. Moreover, the transmission probability associated with the 
$\S$-matrix (\ref{S1}) is given by 
\begin{equation}
\tau(\omega) = |\S_{12}(\sqrt{2m\omega})|^2 = 
\frac{2m\omega (\eta_1-\eta_2)^2 \sin^2 (\theta)}{(2m\omega+\eta_1^2)(2m\omega+\eta_2^2)} \, , 
\qquad \theta \in [0,2\pi)\, , 
\label{tr}
\end{equation}
with 
\begin{equation} 
\eta_i \equiv \eta\, {\rm tan} (\alpha_i)\, , \qquad  \alpha_i \in [-\pi/2, \pi/2)\, , 
\label{tr1}
\end{equation} 
$\left (e^{-2i\alpha_1}, e^{-2i\alpha_2} \right )$ being the eigenvalues of the matrix $\UU$ 
entering the boundary condition (\ref{bc1}). 

The main properties of $\C_1$ and $\C_2$ have been discussed in \cite{L-57, B-86}, whereas  
the non-linear dependence on the chemical potentials has been examined in detail in \cite{MSS-15}. 
Before analysing some $\C_{n\geq3}$, we will face the problem of deriving a generating 
function for the cumulants (\ref{c51},\ref{c52}) and the associated probability distribution.

\bigskip 

\section{Cumulant generating function} 

\medskip 

We show in this section that in spite of the complicated explicit form of the cumulants (\ref{c51},\ref{c52}), 
there exists a relatively simple and compact generating function of $\C(\lambda)$. 
It it instructive to start by extracting the information encoded in (\ref{ec1}-\ref{ec6}) about 
$\C(\lambda)$. Following the pioneering work of Khlus \cite{K-87}, Lesovik and Levitov \cite{L-89, LL-92}, we 
look for a generating function in the form 
\begin{equation}
\C (\lambda ) = \int_0^\infty \frac{\rd \omega}{2\pi}\, 
\ln \left [1+F_{12}(\tau,d_1,d_2)\left (\e^{\ri \lambda f(\tau)}-1\right ) 
+ F_{21}(\tau,d_1,d_2)\left (\e^{-\ri \lambda f(\tau)}-1\right )\right ] \, , 
\label{gf1}
\end{equation} 
where $F_{12}$, $F_{21}$ and $f$ are unknown functions. Using the standard definition 
of generating function 
\begin{equation} 
\C_n = (-\ri \partial_\lambda)^n\, \C(\lambda ) \vert_{{}_{\lambda =0}} 
\label{gft2}
\end{equation}
and the information from the first three cumulants $\C_1$, $\C_2$ and $\C_3$ only, one can 
easily determine $F_{12}$, $F_{21}$ and $f$. The simple result  
\begin{equation}
F_{12} = \frac{1}{2}(c_2 + c_1\sqrt{\tau})\, , \qquad 
F_{21} = \frac{1}{2}(c_2 - c_1\sqrt{\tau})\, , \qquad 
f=\sqrt{\tau}\, . 
\label{gft3}
\end{equation}
leads to the following generating function
\begin{eqnarray}
\C (\lambda ) &=& \int_0^\infty \frac{\rd \omega}{2\pi}\, 
\ln \left [1+\frac{1}{2}(c_2+c_1\sqrt{\tau})\left (\e^{\ri \lambda \sqrt{\tau}}-1\right ) 
+ \frac{1}{2}(c_2-c_1\sqrt{\tau})\left (\e^{-\ri \lambda \sqrt{\tau}}-1\right )\right ]  
\nonumber \\
&=&\int_0^\infty \frac{\rd \omega}{2\pi}\, 
\ln \left \{1+ \ri c_1 \sqrt{\tau}\, \sin (\lambda  \sqrt{\tau}) +c_2 \left [\cos (\lambda  \sqrt{\tau})-1\right ] \right \} \, . 
\label{gf4}
\end{eqnarray}
One can easily check also that (\ref{gf4}) reproduces perfectly the cumulants 
$\C_4$, $\C_5$ and $\C_6$ as well and represents therefore a valid candidate for 
the final result in the case of two leads. The expression (\ref{gf4}) has been reported without 
derivation also by Lesovik and Chtchelkatchev \cite{LC-03}. Our goal below will be 
to generalise (\ref{gf4}) to the multi terminal junction in Fig. \ref{fig1}, thus 
recovering the $N=2$ formula as a special case. In the comments at the end of this section we 
will briefly describe an alternative to (\ref{gf4}), regarding a slightly different setup. 

\bigskip 

\subsection{General result for $N$ terminals} 
\medskip

The argument in what follows is based on the fact that the 
particle transport in our system can be separated 
in statistically independent processes with fixed energy. In fact, excitations with different energies 
propagate in the graph in Fig. \ref{fig1} in a fully independent way, because the only interaction, localised in the vertex,  
leaves the energy invariant (see (\ref{A7})). For this reason one can focus first 
on a single energy channel $\omega$, thus dealing 
with a system with finite degrees of freedom. This fact significantly simplifies the problem and allows to derive 
explicitly the single energy channel cumulant generating function $\C_\omega^i(\lambda )$ 
in the lead $L_i$. The final step is to integrate over all energies, 
\begin{equation} 
\C^i(\lambda ) = \int_0^\infty \frac{\rd \omega}{2\pi}\, \C_\omega^i(\lambda )\, , 
\label{tgf}
\end{equation} 
using at this stage the well known property that the total cumulant of a process, which can be 
decomposed in statistically independent subprocesses, is the sum of the cumulants of each of the latter. 

In order to obtain the particle current of a single energy channel $\omega\geq 0$, we 
modify the integration measure in the general expression of the particle current (\ref{curr1}) according to 
\begin{equation}
\rd k\, \rd p \longmapsto \rd k\, \rd p\, (2\pi)^2\, \delta \left (\frac{k^2}{2m} - \omega \right ) \delta (k-p)\, , 
\label{fe1}
\end{equation} 
which selects the contribution with energy $\omega$. This operation leads to the simple 
time and position independent expression 
\begin{equation}
j_\omega^i = \sum_{l,m=1}^N a^*_l\, \T^i_{lm}(\sqrt{2m\omega})\, a_m \, , 
\label{curr2}
\end{equation}
where $\{a_i, a_i^*\}$ are standard fermionic oscillators: 
\begin{equation}
[a_i\, ,\, a_j^*]_+ = \delta_{ij}\, ,\qquad 
[a_i\, ,\, a_j]_+ = [a^*_i\, ,\, a_j^*]_+ = 0\, , \qquad 
\langle a^*_i a_j\rangle_{\beta,\mu} = \delta_{ij} d_j(\omega)\, . 
\label{alg}
\end{equation}
Now, the generating function of cumulants in the channel $\omega$ and lead $L_i$ is given by 
\begin{equation}
\C_\omega^i(\lambda ) = \ln \langle \e^{\ri \lambda j_w^i} \rangle_{\beta,\mu} \, . 
\label{gf5}
\end{equation} 
The expectation value $\langle \e^{\ri \lambda j_w^i} \rangle_{\beta,\mu}$ 
can be evaluated explicitly. The key points of the computation, which leads to the final result 
\begin{equation}
\langle \e^{\ri \lambda j_w^i} \rangle_{\beta,\mu} = {\rm det} \left [\II + 
\left (\e^{\ri \lambda \T^i(\sqrt{2m\omega})} -\II \right ) \D(\omega) \right ]\, ,   \qquad   
\D(\omega ) \equiv {\rm diag}[d_1(\omega),d_2(\omega),...,d_n(\omega)] \, ,
\label{fe3}
\end{equation} 
are given appendix C. In the lead $L_1$ one finds 
\begin{equation}
\C(\lambda ) = \int_0^\infty \frac{\rd \omega}{2\pi}\, \ln \left \{1+ \ri c_1^{\rm eff} \sqrt{\tau}\, 
\sin \left (\lambda  \sqrt{\tau}\right ) +c_2^{\rm eff}  
\left [\cos \left (\lambda  \sqrt{\tau}\right )-1\right ] \right \} \, . 
\label{gf6}
\end{equation} 
Here 
\begin{equation}
\tau(\omega) = \sum_{i=2}^N \tau_i (\omega)\, , \qquad  \tau_i (\omega) \equiv |\S_{1i}(\sqrt{2m\omega})|^2 \, , 
\label{gf7}
\end{equation}
is the total transmission probability between the lead $L_1$ and the remaining $N-1$ leads $L_i$ and 
$c_{1,2}^{\rm eff}$ are obtained from (\ref{dc}) by the substitution 
\begin{equation}
d_2 (\omega) \longrightarrow d_2^{\rm eff} (\omega) \equiv 
\sum_{i=2}^N\frac{ \tau_i (\omega)}{\tau(\omega)}\, d_i(\omega) \, ,  
\label{gf8}
\end{equation}
which represents an effective distribution where $d_i(\omega)$ with $i\geq 2$ 
are weighted by the ratios $\tau_i(\omega)/\tau(\omega)\in [0,1]$. As expected,  
the expression (\ref{gf6}) reproduces (\ref{gf4}) for $N=2$. 

Summarising, we derived in explicit form the generating function of the cumulants (\ref{c51}, \ref{c52}) for 
the Schr\"odinger junction with $N>2$ leads. 
The novelty with respect to the two terminal case (\ref{gf4}) 
is the effective Fermi distribution (\ref{gf8}), which captures the 
presence of all $N-1>1$ reservoirs.

\bigskip 
\subsection{Scale invariant limit} 
\medskip 

The $\omega$-integration in (\ref{gf4},\ref{gf6}) 
with general $\S$-matrix of the form (\ref{S1}) cannot 
be performed in a closed analytic form. For this reason it is instructive to select among (\ref{S1}) 
the {\it scale-invariant} matrices, which incorporate the universal transport properties of the system \cite{BDV-15} 
while being simple enough to be analysed explicitly. The scale invariant (critical) elements 
in the family (\ref{S1}) are fully classified \cite{Calabrese:2011ru} and belong to the orbits 
\begin{equation} 
\{{\cal U}\, \S_d\, {\cal U}^* \; :\; {\cal U} \in U(N),\; \S_d = {\rm diag}(\pm 1,\, \pm 1,\, ...\, ,\, \pm 1)\} 
\label{si1}
\end{equation} 
of the adjoint action of the unitary group $U(N)$ on the diagonal matrices $\S_d$. As expected, 
the critical $\S$-matrices are $\omega$-independent, which in the case $\beta_1=\beta_2\equiv \beta$ 
allows to compute the integrals in (\ref{gf4},\ref{gf6}) explicitly. In fact, introducing the variables
\begin{equation} 
\gamma_j\equiv\e^{-\beta\mu_j}\, ,\qquad 
\Lambda(\lambda ) \equiv \ri \sqrt{\tau}(\gamma_2-\gamma_1)\sin(\lambda\sqrt{\tau})
	+(\gamma_2+\gamma_1)\cos(\lambda\sqrt{\tau})\, ,
\label{si2}
\end{equation} 
where $\tau=|\S_{12}|^2$ now is constant, one gets in the case $N=2$  
\begin{equation}
\begin{split}
\C(\lambda) = \frac{1}{2\pi \beta}\Bigg\{
&\Li_2\left (-\gamma_1^{-1}\right ) + \Li_2\left (-\gamma_2^{-1}\right )
\\
-&\Li_2\left[\frac{-2}{\Lambda(\lambda) -
\sqrt{\Lambda^2(\lambda)-4\gamma_1\gamma_2}}\right]
- \Li_2\left[\frac{-2}{\Lambda(\lambda) +
\sqrt{\Lambda^2(\lambda)-4\gamma_1\gamma_2}}\right] \Bigg\}\, , 
\label{si3}
\end{split}
\end{equation} 
$\Li_2$ being the dilogarithm function. 

The results of \cite{MSS-15} suggest to investigate (\ref{si3}) as a function of 
\begin{equation}
\mu_\pm = (\mu_1 \pm \mu_2)/2 \, , 
\label{mupm}
\end{equation} 
$\mu_+ \in \RR$ playing the role of control parameter. For $\mu_+=0$ 
the expression (\ref{si3}) greatly simplifies 
in the low temperature limit $\beta \to \infty$. In fact, one finds 
\begin{equation}
\lim_{\beta \to \infty} \C (\lambda )\vert_{{}_{\mu_+=0}}  = 
\frac{|\mu_-|}{2\pi} \ln \left [\cos (\lambda \sqrt{\tau})
+ \ri\, \varepsilon (\mu_-)\sqrt{\tau} \sin (\lambda \sqrt{\tau})\right ] \, ,  
\label{si4}
\end{equation} 
$\varepsilon$ being the sign function. The result (\ref{si4}) was derived for $\mu_->0$ independently 
by Levitov and Lesovik \cite{LL-92, LS-14} and provides therefore a valuable check 
on (\ref{si3}). For the first few cumulants one gets from (\ref{si4}) 
\begin{equation}
\C_1 = \frac{\tau \mu_-}{2\pi} \, ,\qquad 
\C_2 = \frac{\tau (1-\tau)|\mu_-|}{2\pi}\, ,\qquad 
\C_3 = \frac{\tau^2(\tau-1) \mu_-}{\pi}\, ,\qquad 
\C_4 = \frac{ \tau^2(\tau-1)(1-3\tau) |\mu_-|}{\pi}\, .  
\label{si5}
\end{equation} 
From (\ref{si4}) one infers that for any $n\geq 0$ 
\begin{equation}
\C_{2n+1}\propto \C_1\, , \qquad \mu_+=0\, , \quad \beta \to \infty\, , 
\label{prop}
\end{equation}
which provides an interesting signature for experiments. In fact, 
experimental evidence for the linear dependence of $\C_3$ on 
the current $\C_1$ in this regime has been reported in \cite{BGSLR-05}. 

It is instructive to derive the charge transferred through the junction in the time interval $[0,T]$. 
Since the LB state is stationary, one has in general 
\begin{equation}
Q_T = \int_0^T \rd t\, \langle j(t,x,i)\rangle_{\beta,\mu} = T \C_1\, .   
\label{si6}
\end{equation} 
Restoring the electric charge according to $j\longmapsto e j$ and $\mu_-=eV$, where $V$ is 
the applied voltage, one gets from (\ref{si5})  
\begin{equation}
Q_T = \frac{ e^2\tau}{2\pi} VT = \frac{ e_{\rm eff}^2}{2\pi} VT\, ,  \qquad e_{\rm eff}\equiv e \sqrt {\tau}\, ,
\label{si7}
\end{equation} 
where the effective charge $e_{\rm eff}$ has been introduced. 
We see that switching on the defect ($\tau < 1$) causes a finite 
renormalisation of the charge $e \longmapsto e_{\rm eff}$ 
with respect to the case in which the defect is absent ($\tau=1$). 
This purely quantum phenomenon is induced by the non-trivial reflection probability $(1-\tau)$ from 
the defect. In this respect the appearance of the effective charge $e_{\rm eff}$ in the 
probability distribution (\ref{Mo7}), reconstructed in section 4 below, is not surprising. It is worth mentioning 
that the same charge renormalisation effect has been observed in \cite{LL-92, LC-03} as well. 

\begin{figure}[h]
\begin{center}
\begin{picture}(280,70)(80,25) 
\includegraphics[scale=0.50]{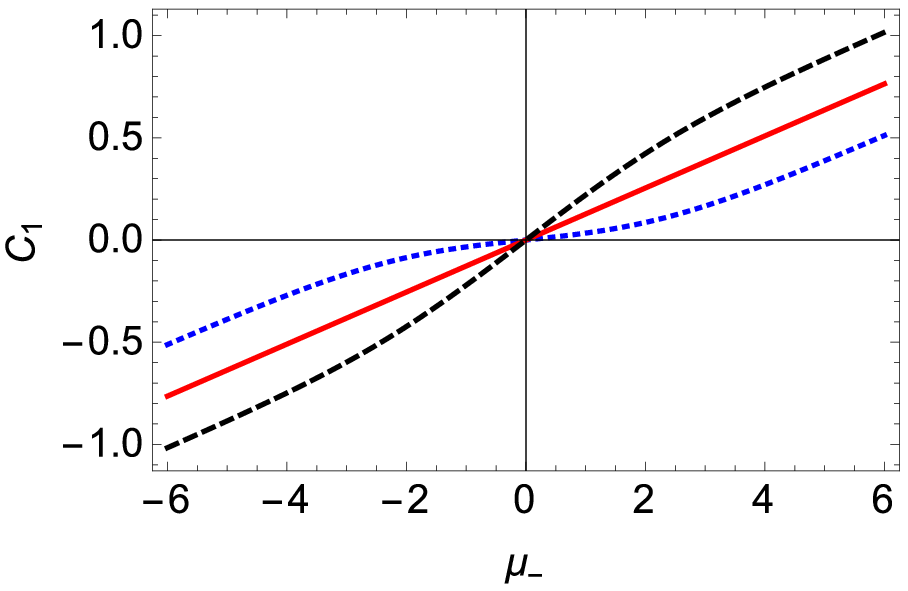} 
\hskip 0.5 truecm
\includegraphics[scale=0.50]{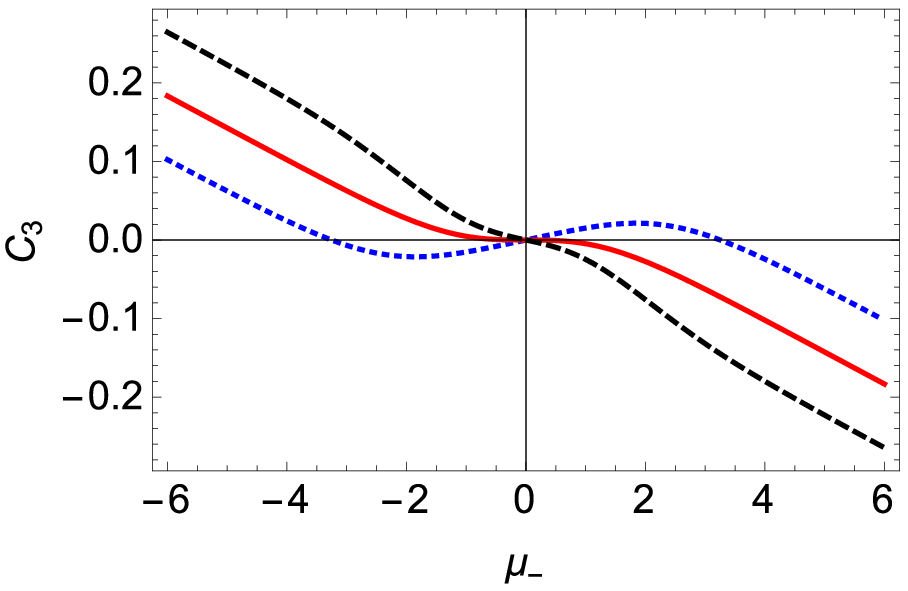}
\hskip 0.5 truecm
\includegraphics[scale=0.50]{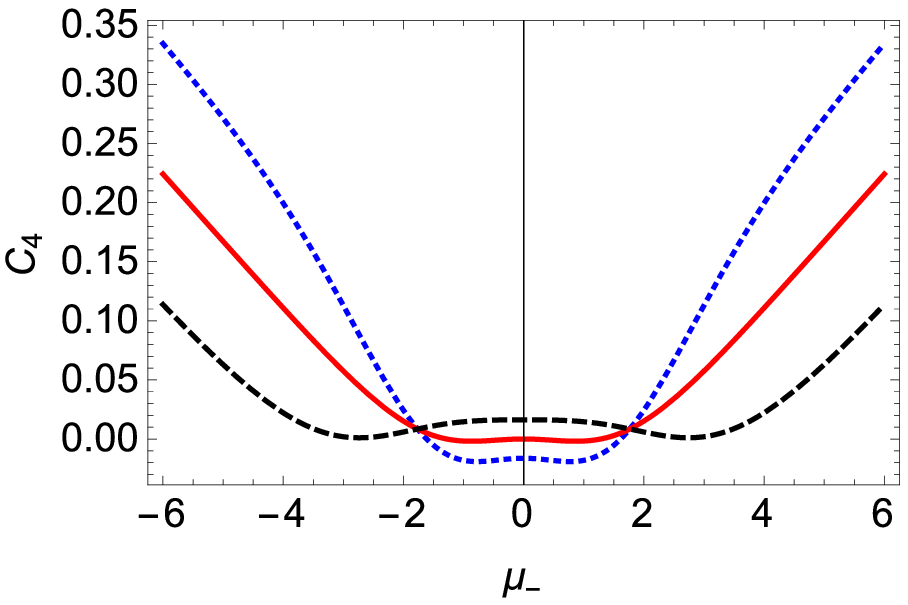}
\end{picture} 
\end{center}
\caption{The cumulants $\C_1$ (left), $\C_3$ (center) and $\C_4$ (right) (in units of $1/\beta$) in function 
of $\mu_-$ for $\beta=1$, $\tau=0.8$ and $\mu_+=-2$ (blue dotted line), $\mu_+=0$ (red continuous line) 
and $\mu_+=2$ (black dashed line).}
\label{fig2}
\end{figure}

Relevant non-linear effects show up \cite{MSS-15} for $\mu_+\not=0$. Consider for instance 
the current, which can be written in the form
\begin{equation} 
\C_1=  \frac{\tau \mu_-}{2\pi}  + 
\frac{\tau}{2\pi \beta} 
\ln \left [\frac{\e^{\beta \mup}+\e^{-\beta \mum}}{1+\e^{\beta(\mup-\mum)}}\right ] \, . 
\label{si8}
\end{equation} 
The second term vanishes for $\mu_+=0$, but is nontrivial otherwise and captures the 
nonlinear behavior close to the origin $\mu_-=0$, shown in the left panel of Fig. \ref{fig2}. 
Remarkably enough, such nonlinearity has been experimentally observed in the 
junctions studied in \cite{BJS-95}. The reduction and enhancement \cite{MSS-15} 
of the noise power $\C_2$ for $\mu_+<0$ and $\mu_+>0$ are also consequences 
of the nonlinearity in $\mu_-$. 

It follows from (\ref{si3}) that for $\mu_+\not=0$ the cumulants $\C_3$ and 
$\C_4$ are respectively odd and even nonlinear functions of $\mu_-$. For the slopes 
of the asymptotes at $\mu_-=\pm \infty$ one finds 
\begin{equation}
\lim_{\mu_-\to \pm \infty} \frac{\C_3}{\mu_-} = \frac{\tau^2(\tau-1)}{\pi}\, ,\qquad 
\lim_{\mu_-\to \pm \infty} \frac{\C_4}{\mu_-} = \pm \frac{ \tau^2(\tau-1)(1-3\tau)}{\pi}\, , 
\label{asymp}
\end{equation} 
which depend on $\tau$ but not on $\beta$ and $\mu_+$. The central and the right panel 
of Fig. \ref{fig2} illustrate the behavior of $\C_3$ and $\C_4$ for three different values of 
$\mu_+$.

\bigskip 

\subsection{Comments} 

\medskip 

The form of the current correlations (\ref{w1},\ref{c1}) clearly depends on the non-equilibrium system under 
consideration. In the present investigation the cumulants (\ref{c51}, \ref{c52}) refer to the family of point-like 
defects defined by the boundary condition (\ref{bc1}) or equivalently, by the scattering matrix (\ref{S1}). 
In the paper \cite{LLL-96} Levitov, Lee and Lesovik (LLL) investigated a different 
setup. First of all, instead of the boundary condition (\ref{bc1}), which  
fixes the self-adjoint extension of the operator $-\partial_x^2$ in our framework, these authors 
introduced in (\ref{eqm1}) a minimal coupling 
\begin{equation}
\ri \partial_x \longrightarrow \ri \partial_x + A(x)\, . 
\label{minc1}
\end{equation}  
The physical idea, inspiring this 
modification, is to implement a kind of quantum galvanometer, based 
on the interaction of the current (\ref{curr1}) with the classical potential $A(x)$. 
In addition, instead of the zero frequency limit (\ref{c4}) of the current expectation values (\ref{c3}), 
LLL focus on 
\begin{equation}
\int_0^T \rd t_1\cdots \int_0^T \rd t_n\, \langle j(t_1,x_1,i)\cdots j(t_1,x_1,i)\rangle_{\beta,\mu}  
\label{LL0}
\end{equation} 
at $x_1=\cdots =x_n=0$. In the special case 
\begin{equation}
A(x) \sim \lambda\, \delta(x)
\label{minc2}
\end{equation}
with a coupling proportional to the counting parameter $\lambda$, LLL got for 
sufficiently large $T$ \cite{LLL-96} (see also \cite{Sch-07}, \cite{LS-14} and 
the contributions to \cite{N-03})
\begin{eqnarray}
\C_{L} (\lambda ) &=& \int_0^\infty \frac{\rd \omega}{2\pi}\, 
\ln \left [1+\frac{1}{2}(c_2+\tau c_1)\left (\e^{\ri \lambda}-1\right ) 
+ \frac{1}{2}(c_2-\tau c_1)\left (\e^{-\ri \lambda}-1\right )\right ]  
\nonumber \\
&=&\int_0^\infty \frac{\rd \omega}{2\pi}\, 
\ln \left \{1+ \ri c_1 \tau\, \sin (\lambda) + \tau c_2 \left [\cos (\lambda ) -1\right ]\right \} \, . 
\label{LL1}
\end{eqnarray} 

As observed already in \cite{LC-03}, the difference between (\ref{gf4}) and (\ref{LL1}) is not surprising, because 
these results concern two inequivalent settings. Notice in this respect the twofold role 
of $\lambda$ in \cite{LLL-96}  as a coupling constant (\ref{minc2}) and a counting parameter, whereas 
the system (\ref{eqm1},\ref{bc1}) and the cumulants (\ref{c51},\ref{c52}) 
are $\lambda$-independent and $\lambda$ enters in (\ref{gf4}, \ref{gf6}) 
only as an auxiliary parameter in the spirit of conventional generating functions. 
Nevertheless, comparing (\ref{gf4}) and (\ref{LL1}), we observe that these results coincide 
when the defect in the junction disconnects the leads ($\tau=0$), or when 
the defect is absent ($\tau=1$).  Moreover, quite remarkably the first two 
cumulants generated by (\ref{gf4}) and (\ref{LL1}) coincide for any $\tau$. 
Therefore the transferred charge (\ref{si6},\ref{si7}) and the zero frequency quantum noise power 
$\C_2$ coincide in both schemes. For distinguishing the two settings, one should 
concentrate on the higher cumulants $\C_{n\geq 3}$. It turns out that they 
have the same monomials in $c_1$ and $c_2$, but with different 
powers of $\tau$. One has for instance 
\begin{eqnarray} 
\C_{L, 3} &=& \int_0^\infty \frac{\rd \omega}{2\pi}\, \tau c_1(1-3\tau c_2+2\tau c_1^2) \, ,
\label{LL2}\\\
\C_{L, 4} &=& \int_0^\infty \frac{\rd \omega}{2\pi}\, \tau[c_2 -3\tau c_2^2+12\tau^2 c_1^2c_2 - 
2\tau c_1^2(2+3\tau^2 c_1^2)]\, , 
\label{LL3}
\end{eqnarray}
to be compared with (\ref{ec3},\ref{ec4}). The difference between $\C_3$ and $\C_{L,3}$ has been 
studied in detail in \cite{LC-03}. It has been argued there that 
both $\C_3$ and $\C_{L,3}$ are in principle observable, but in different experimental setups. 

It is natural to expect that the form (\ref{LL1}) of $\C_L$ depends on the specific choice (\ref{minc2})  
of the external potential, which implements the effective description of the charge detector. It has been 
argued in \cite{NK-03} that localising $A(x)$ at $x=0$ 
minimises the disturbance of the system due to the measuring 
device. The localisation of $A(x)$ in one point can be achieved however 
in many different ways, using general linear combinations of the delta function and its derivatives. 
The study of the freedom associated with $A(x)$ is 
beyond the scope of the present paper, which is focussed on  
the properties of the cumulants (\ref{c51},\ref{c52}), uniquely defined 
by the connected current correlation functions 
(\ref{c1}) in the quantum boundary value problem (\ref{eqm1},\ref{bc1}). 
In the next section we recover the probability distribution, associated with these cumulants, 
and provide a microscopic physical interpretation for it.

\bigskip 

\section{Moments and probability distribution}
\medskip 

Since the cumulants (\ref{c51},\ref{c52}) concern a quantum field theory system with 
unitary time evolution, one can expect that they correspond to a well defined probability distribution. 
In order to show that this is indeed the case, we reconstruct below this distribution from its moments, 
following a standard procedure in probability theory \cite{ST-70}. In our case 
the moment generating function at fixed energy $\omega$ for the two-lead 
junction can be extracted from equation (\ref{C4}) in appendix C. One has 
\begin{equation}
\chi_\omega (\lambda) = 1+ \ri c_1\sqrt{\tau}\, \sin \left (\lambda \sqrt{\tau }\right )+
c_2 \left [\cos \left (\lambda \sqrt{\tau} \right )-1\right ]\, , 
\label{Mo0}
\end{equation} 
where $c_i(\omega)$ and $\tau(\omega)$ are given by (\ref{dc}-\ref{tr1}). The moments $\{m_n\, :\, n=0,1,...\}$ are inferred  
from the expansion 
\begin{equation} 
\chi_\omega (\lambda) = \sum_{n=0}^\infty \frac{(\ri \lambda)^n}{n!}\, m_n
\label{Mo1}
\end{equation}
and have the simple general form 
\begin{equation} 
m_n  = 
\begin{cases} 
\,1 \, , & n=0\, , \\
c_1 \tau^{k}\, , & n=2k-1\, ,\quad \; k=1,2,...\, , \\ 
c_2 \tau^{k}\, , & n=2k\, ,\quad \qquad k=1,2,...\, . \\ 
\end{cases} 
\label{Mo2}
\end{equation} 

One can verify that (\ref{Mo2}) and the cumulants at energy $\omega$, given by the integrands of 
(\ref{c51}, \ref{c52}) (see also (\ref{ec1}-\ref{ec6})), satisfy the conventional relations between 
moments and cumulants. The nice surprise is that for generic $n$ the moment $m_n$ 
is much simpler then the corresponding cumulant. This fact represents a great technical 
advantage for solving the moment problem, namely for determining 
a probability distribution $\varphi (\xi)$ such that 
\begin{equation} 
m_n = \int_{\cal D} \rd \xi\, \xi^n \varphi (\xi)\, . 
\label{Mo3}
\end{equation} 
There exist \cite{ST-70} three possible choices for the domain $\cal D$: the whole line ${\cal D}=\RR$, 
the half line ${\cal D}= \RR_+$ and a compact interval ${\cal D}=[a,b]$. 
A necessary and sufficient condition for the existence of 
$\varphi$ on $\RR$ is \cite{ST-70} the non-negativity of the Hankel determinants  
\begin{equation} 
\HH_n \equiv 
\left | \begin{array}{cccccccc}
m_0&m_1&\cdots &&  m_n\\
m_1&m_2& \cdots & &m_{n+1}\\
\vdots &\vdots & \vdots&& \vdots \\
m_n&m_{n+1}& \cdots & &m_{2n}\\
\end{array}\right | \geq 0\, . 
\label{Mo4}
\end{equation} 
{}From (\ref{Mo3}) one gets 
\begin{equation}
\HH_0 = 1\, , \qquad \HH_1=\tau(c_2-c^2_1\tau)\, , 
\qquad \HH_2=\tau^3(1-c_2)(c_2^2-c_1^2\tau)\, , \qquad \HH_{n\geq 3} = 0\, .
\label{Mo5}
\end{equation} 
Using the explicit form of $c_i$ and that $\tau \in [0,1]$, 
one can show that $\HH_2$ and $\HH_3$ are non-negative. Therefore the probability 
distribution $\varphi$ on $\RR$ exists. Combining the results of \cite{ST-70} with 
\begin{equation} 
\HH_2^\prime \equiv 
\left | \begin{array}{cccccccc}
m_1&m_2\\
m_2&m_3\\
\end{array}\right | = \tau^2 (c_1\tau -c_2) \leq 0\, ,  
\label{Mo4bis}
\end{equation} 
one concludes that the domains $\RR_+$ and $[a,b]$ are excluded and one il left therefore 
with the so called Hamburger moment problem ${\cal D}=\RR$. Moreover, since $\HH_{n\geq 3} = 0$ the  
general theory \cite{ST-70} implies that $\varphi$ 
is localised in three different points on the $\xi$-line. In fact, employing (\ref{Mo0}-\ref{Mo3}), 
one finds 
\begin{equation}
\begin{split}
\varphi (\xi) &= \int_{-\infty}^\infty \frac{\rd \lambda}{2\pi}\, \e^{-\ri \lambda \xi}\, \chi_\omega (\lambda ) \\
&=\frac{1}{2}(c_2-c_1\sqrt {\tau})\delta(\xi +e\sqrt {\tau})+(1-c_2) \delta(\xi) + 
\frac{1}{2}(c_2+c_1\sqrt {\tau})\delta(\xi -e \sqrt {\tau}) \, , 
\label{Mo7}
\end{split}
\end{equation} 
where the charge $e$ has been restored for clarifying the physical interpretation of $\varphi$. 
The normalisation condition 
\begin{equation}
\int_{-\infty}^\infty \rd \xi\, \varphi (\xi) = 1  
\label{Mo8}
\end{equation}
is easily verified. One can show in addition that the coefficients of the delta 
distributions in (\ref{Mo7}) take value in the interval $[0,1]$ and have therefore 
a direct probabilistic interpretation. In analysing this fundamental point it is useful to 
distinguish the two cases $\tau\not=0$ and $\tau=0$. The basic microscopic 
process, which takes place in the system for $\tau\not=0$, is the emission 
of a particle with energy $\omega$ from a reservoir 
$R_i$ and its absorption from $R_j$. Let us denote by $p_{ij}$ the relative 
probability. Recalling that $\varphi$ concerns the lead $L_1$, the three terms in 
(\ref{Mo7}) correspond to the three elementary processes of emission and absorption 
relative to the reservoir $R_1$, namely: 

(i) $p_{12}=\frac{1}{2}(c_2-c_1\sqrt {\tau})$ is the probability for a particle to be 
emitted from $R_1$ and absorbed by $R_2$; 

(ii) $p_{11}=(1-c_2)$ is the probability for a particle to be emitted from and absorbed 
by $R_1$; 

(iii) $p_{21}=\frac{1}{2}(c_2+c_1\sqrt {\tau})$ is the probability for a particle 
to be emitted from $R_2$ and absorbed by $R_1$.  

The variation $\xi$ of the charge in $L_1$, involved in these processes, is 
fixed by the support of the delta functions in (\ref{Mo7}) and is $\xi = e_{\rm eff}$, 
$\xi=0$ and $\xi=- e_{\rm eff}$ respectively, $e_{\rm eff}$ being defined in (\ref{si7}).  
These values of $\xi$ are consistent with the physical interpretation of (i)-(iii). It is 
worth stressing that $p_{11}$ is $\tau$-independent as intuitively expected. 

For $\tau=0$ the two leads are disconnected and 
one expects therefore that $p_{12} = p_{21}=0$ and $p_{11} =1$. 
This simple physical observation is confirmed by the fact that the three 
terms in (\ref{Mo7}) collapse in one, namely
\begin{equation}
\varphi (\xi)\vert_{\tau=0} = \delta(\xi)\, .  
\label{Mo7b} 
\end{equation} 
The property (\ref{Mo7b}) implies also that as functions of $\tau$ the probabilities $p_{ij}$ 
are discontinuous at $\tau=0$, when the system divides in two parts. 

The probability distribution $\varphi$ in $L_2$ is obtained by implementing $d_1 \leftrightarrow d_2$, 
or equivalently (see (\ref{dc})) by the substitutions $c_1 \to -c_1$ and $c_2 \to c_2$ in (\ref{Mo7}). 
As expected, under this operation one has $p_{12} \leftrightarrow p_{21}$. We deduce moreover  
that $p_{22}=p_{11}$, which completes the picture of the two lead junction at microscopic level. 

It is instructive to study the behavior of the probabilities $p_{ij}$ because they 
provide fundamental information about the elementary processes in the system 
and uniquely fix the probability distribution, thus determining all 
moments and cumulants. Since 
\begin{equation}
p_{11}+p_{12}+p_{21}=1\, ,  
\label{b1}
\end{equation}
it is enough to focus on the pair $\{p_{12}, p_{21}\}$. 

\begin{figure}[h]
\begin{center}
\begin{picture}(140,70)(80,25) 
\includegraphics[scale=0.50]{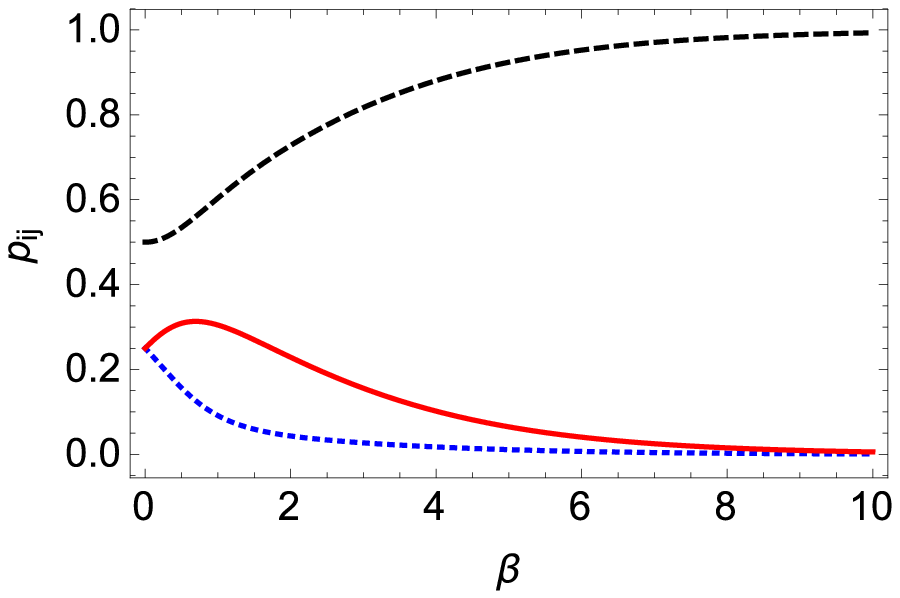} 
\hskip 1 truecm
\includegraphics[scale=0.50]{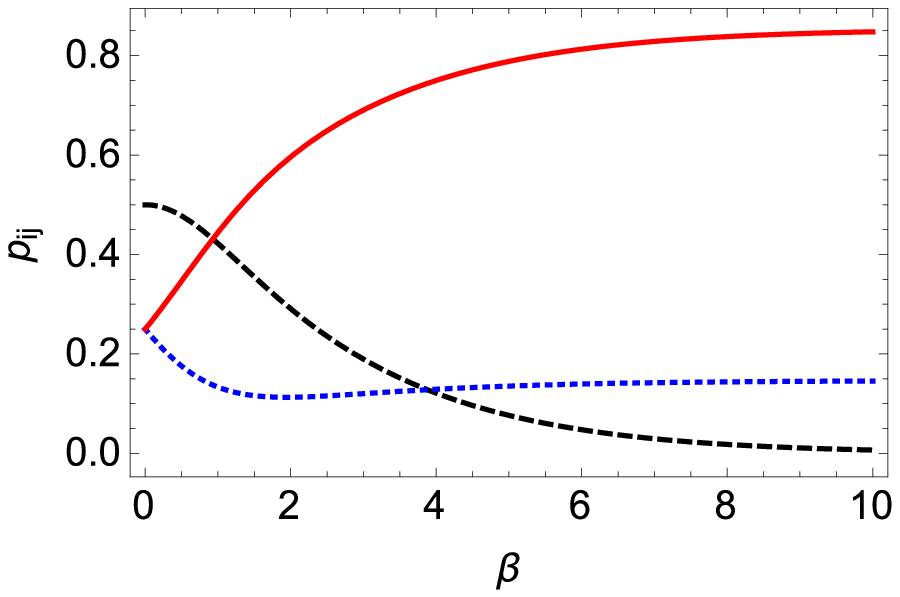}
\end{picture} 
\end{center}
\caption{Temperature dependence of $p_{11}$ (black dashed line), $p_{12}$ 
(blue dotted line) and $p_{21}$  (red continuous line)
for $\mu_+$ outside (left) and inside (right) of $[\omega-\mu_-\, ,\, \omega+\mu_-]$ and $\tau=0.5$.}
\label{fig3}
\end{figure}

{}For describing the low temperature limit $\beta_1=\beta_2 \equiv \beta \to \infty$ it is convenient 
to adopt the variables $\mu_\pm$ defined by (\ref{mupm}),  
assuming without loss of generality that $\mu_-\geq 0$. Then one finds  
\begin{equation} 
\lim_{\beta \to \infty} p_{12}= 
\begin{cases} 
0 \, , & \qquad  \mu_+ < \omega-\mu_-\, , \\
(1-\sqrt{\tau})/2\, , & \qquad \omega -\mu_- < \mu_+ < \omega + \mu_-\, , \\ 
0 \, , & \qquad  \mu_+ > \omega+\mu_-\, , \\
\end{cases} 
\label{b2}
\end{equation} 
and 
\begin{equation} 
\lim_{\beta \to \infty} p_{21}= 
\begin{cases} 
0 \, , & \qquad  \mu_+ < \omega-\mu_-\, , \\
(1+\sqrt{\tau})/2\, , & \qquad \omega -\mu_- < \mu_+ < \omega + \mu_-\, ,\\ 
0 \, , & \qquad  \mu_+ > \omega+\mu_-\, . \\
\end{cases} 
\label{b3}
\end{equation} 
At the boundary points $\mu_+ = \omega \pm \mu_-$ one has instead
\begin{eqnarray}
\lim_{\beta \to \infty} p_{12}&=&(1-\sqrt{\tau})/4\, ,\qquad 
\lim_{\beta \to \infty} p_{21}=(1+\sqrt{\tau})/4\, ,\qquad \mu_- \not= 0 \, , \nonumber \\
\lim_{\beta \to \infty} p_{12}&=& 
\lim_{\beta \to \infty} p_{21}= 1/4\, ,\qquad \mu_- = 0 \, .  
\label{b23}
\end{eqnarray}
We conclude that at low temperatures the process of emission and absorption 
from the same reservoir is favored for $\mu_+$ outside the interval $[\omega-\mu_-\, ,\, \omega+\mu_-]$, 
which is illustrated by the left panel of Fig. \ref{fig3}.  
The probabilities $p_{21}$ and $p_{12}$ are instead dominating 
if $\mu_+\in (\omega-\mu_-\, ,\, \omega+\mu_-)$, 
as shown in the right panel of the same figure. Finally, equation (\ref{b23}) implies that in the low temperature limit 
$p_{11}=1/2$ for $\mu_+=\omega \pm \mu_-$. 

At high temperatures one has 
\begin{equation} 
\lim_{\beta\to 0} p_{12} = \lim_{\beta \to 0} p_{21} =1/4\, ,  
\label{b4}
\end{equation} 
which is manifest in Fig. \ref{fig3}. 

\begin{figure}[h]
\begin{center}
\begin{picture}(280,70)(80,25) 
\includegraphics[scale=0.50]{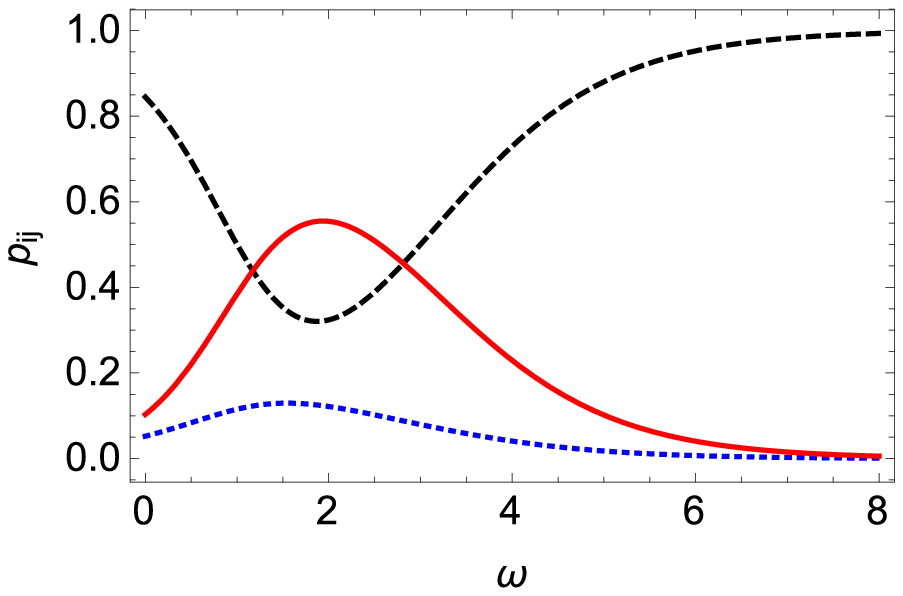} 
\hskip 0.5 truecm
\includegraphics[scale=0.50]{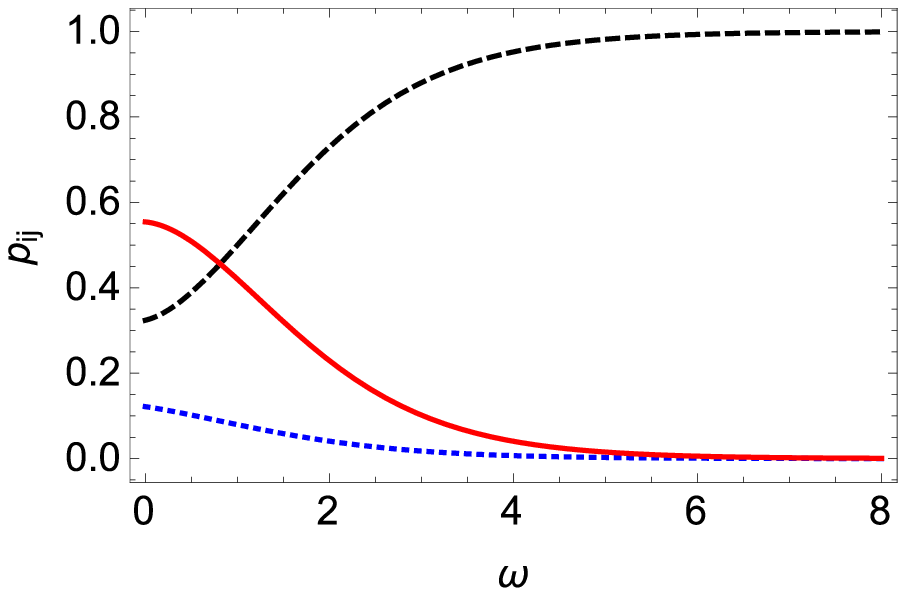}
\hskip 0.5 truecm
\includegraphics[scale=0.50]{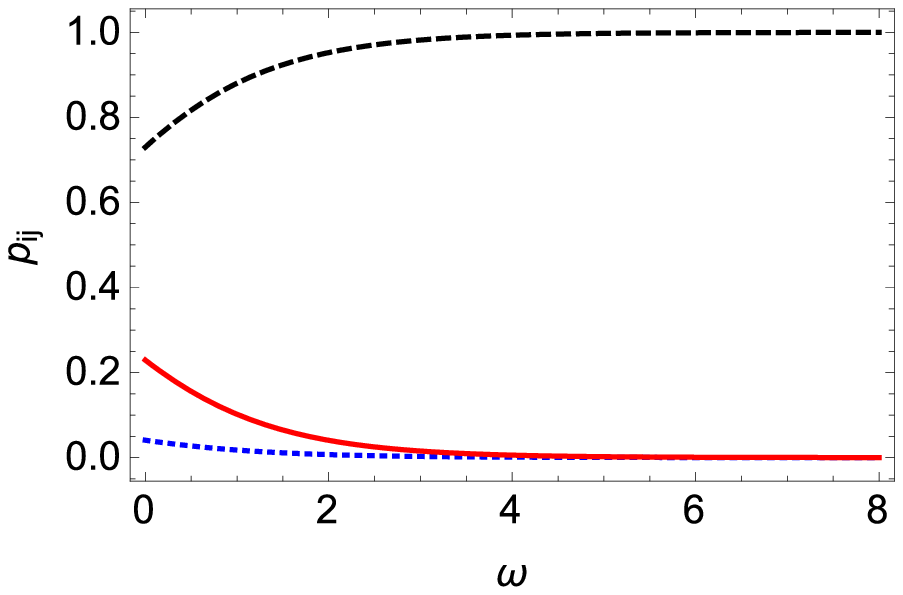}
\end{picture} 
\end{center}
\caption{Energy dependence of $p_{11}$ (black dashed line), $p_{12}$ 
(blue dotted line) and $p_{21}$  (red continuous line)
for $\mu_+>0$ (left),  $\mu_+=0$ (center) and $\mu_+<0$(right).}
\label{fig4}
\end{figure}

At high energies one finds  
\begin{equation} 
\lim_{\omega\to \infty} p_{12} = \lim_{\omega\to \infty} p_{21} =0\, ,
\label{b5}
\end{equation} 
showing that in this regime the probability of emission of a particle from one reservoir and 
its absorption from the other one is negligible. This feature is illustrated by the blue and red 
curves in the plots of Fig. \ref{fig4} for constant $\tau=0.5$. 

Summarising, the solution of the moment problem at hand is a superposition of three delta functions. The  
coefficients are positive and have direct physical interpretation, representing the 
probabilities of emission and absorption of particles from the heat reservoirs of the system. 

We conclude with the following two observations. First, in terms of 
the basic probabilities $\{p_{ij}, \tau \}$, the cumulant 
generating function (\ref{gf4}) takes the form 
\begin{eqnarray}
\C (\lambda ) &=& \int_0^\infty \frac{\rd \omega}{2\pi}\, 
\ln \left \{1+ \ri (p_{21}-p_{12})\, \sin (\lambda  \sqrt{\tau}) +
(p_{21}+p_{12})\left [\cos (\lambda  \sqrt{\tau})-1\right ] \right \} 
\nonumber \\ 
&=&\int_0^\infty \frac{\rd \omega}{2\pi}\, 
\ln \left [p_{11}+ p_{21}\, \e^{\ri \lambda \sqrt{\tau}} + p_{12}\, \e^{-\ri \lambda \sqrt{\tau}} \right ] \, . 
\label{gf4bis}
\end{eqnarray}
Second, the distribution (\ref{Mo7}) has a straightforward extension 
to the multi terminal system $N\geq 3$. In this case the probability distribution $\varphi (\xi)$ in the lead 
$L_1$ is obtained from (\ref{Mo7}) by replacing $c_i$ by the effective quantities $c_i^{\rm eff}$ 
according to (\ref{gf8}) and fixing $\tau$ by (\ref{gf7}). Now $p_{12}$ is the probability of 
the emission of a particle from $R_1$ and absorption by any of the reservoirs 
$\{R_2, R_3,...,R_N\}$, whereas $p_{21}$ gives the probability of the inverse process. 

\bigskip 

\section{Outlook and conclusions} 

\medskip

We developed a quantum field theory approach for the derivation of the $n$-point particle 
current correlation functions for the system in Fig. \ref{fig1}, modelling 
a quantum wire junction away from equilibrium. The system has the geometry of a star 
graph with free particle propagation along the leads and a defect interaction 
localised in the vertex. The new achievements in this framework are:  
\medskip 

(a) the closed and exact form (\ref{c51}-\ref{c52}) of the cumulants $\C_n$ for generic $n$;  

(b) the cumulant generating function (\ref{gf6}-\ref{gf8}) in the $N$-terminal case; 

(c) the cumulant generating function (\ref{si3}) in the scale invariant limit; 

(d) the moments (\ref{Mo2}) and the associated probability distribution (\ref{Mo7}); 

(e) the exact emission-absorption probabilities $p_{ij}$. 
\medskip 

These results clearly indicate that the probabilities $p_{ij}$, which describe 
the fundamental microscopic processes in the system and are the final goal of our investigation,  
represent the core of the quantum transport problem in consideration. In fact, our analysis shows
that $p_{ij}$, supplemented by the defect transmission $\tau$, 
are the building blocks of the probability distribution and uniquely determine all moments and cumulants. 
We established the explicit form of $p_{ij}$ and analysed in detail their dependence 
on the heat bath parameters $\{\beta_i, \mu_i\}$. 

The field theoretical framework, developed in this paper, is universal and can be applied to other systems as well. 
It will be interesting for instance to extend the above results to Majorana fermions, which attract  
much attention \cite{M-1,M-2} with their remarkable physical 
properties and potential applications in topological quantum 
information \cite{K-01}. In this context the quantum transport of Majorana fermions along the edge of 
a topological superconductor represents a fascinating problem, which can be faced by the above 
methods.

\medskip 

%\vfill\eject 

\appendix
\bigskip 
\section{Correlation functions in the LB state} 

\medskip 

In the computation of the current correlators (\ref{w1}) we need the $2n$-point functions 
\begin{equation}
\langle a^*_{l_1}(k_1) a_{m_1}(p_1)\cdots a^*_{l_n}(k_n) a_{m_n}(p_n)\rangle_{\beta,\mu} 
\label{A1}
\end{equation} 
for {\it positive} momenta. In this case 
\begin{equation}
\langle a^*_{l_1}(k_1) a_{m_1}(p_1)\cdots a^*_{l_n}(k_n) a_{m_n}(p_n)\rangle_{\beta,\mu} = 
\frac{1}{Z} {\rm Tr} \left [\e^{-K} a^*_{l_1}(k_1) a_{m_1}(p_1)\cdots a^*_{l_n}(k_n) a_{m_n}(p_n)\right ]\, , 
\label{A2}
\end{equation} 
where 
\begin{equation}
K= \int_0^\infty \frac{\rd k}{2\pi} \sum_{i=1}^N \beta_i \left [\omega(k) - \mu_i\right ]a^*_i(k) a_i(k) \, , 
\quad Z = {\rm Tr} \left (\e^{-K}\right )\, . 
\label{A3}
\end{equation}
{}For the two-point functions one gets ($k>0,p>0$)
\begin{eqnarray} 
\langle a_l^*(k)a_m(p)\rangle_{\beta, \mu} &=& 
2\pi \delta (k-p)\delta_{lm}d_l[\omega(k)] \equiv \Delta_{lm}(k,p)\, , 
\label{A4} \\
\langle a_m(p) a_l^*(k)\rangle_{\beta, \mu} &=& 
2\pi \delta (k-p)\delta_{lm}\{1-d_l[\omega(k)]\} \equiv \widetilde{\Delta}_{lm}(k,p)\, , 
\label{A5}
\end{eqnarray} 
where $d_l(\omega)$ is the Fermi distribution (\ref{c11}) of the reservoir $R_l$. 

The $2n$-point function can be expressed in terms of (\ref{A4},\ref{A5}) as a determinant. For  
$k_i>0,p_i>0$ one has:
\begin{eqnarray}
\langle a^*_{l_1}(k_1) a_{m_1}(p_1)\cdots a^*_{l_n}(k_n) a_{m_n}(p_n)\rangle_{\beta,\mu} = 
\qquad \qquad \quad  \nonumber \\
\left | \begin{array}{cccccccc}
\Delta_{l_1m_1}(k_1,p_1)&\Delta_{l_1m_2}(k_1,p_2)&\cdots &&  \Delta_{l_1m_n}(k_1,p_n)\\
-{\widetilde{\Delta}}_{l_2m_1}(k_2,p_1)&\Delta_{l_2m_2}(k_2,p_2) & \cdots & &\Delta_{l_2m_n}(k_2,p_n) \\
\vdots &\vdots & \vdots&& \vdots \\
-{\widetilde{\Delta}}_{l_nm_1}(k_n,p_1)&-{\widetilde{\Delta}}_{l_nm_2}(k_n,p_2) & \cdots & &\Delta_{l_nm_n}(k_n,p_n)\\
\end{array}\right | \, .
\label{A6}
\end{eqnarray} 
The validity of \eqref{A6} can be proven by induction. For $n=1$ the determinant correctly 
reproduces (\ref{A4}). Using the definition (\ref{A2},\ref{A3}) one can show that for $n\geq2$ 
\begin{equation}
\langle a^*_{l_1}(k_1) a_{m_1}(p_1)\cdots a^*_{l_n}(k_n) a_{m_n}(p_n)\rangle_{\beta,\mu} 
= \sum_{k=1}^n \langle a^*_{l_1}(k_1) a_{m_k}(p_k)\rangle_{\beta,\mu}
\langle \, \cdots \,  \rangle'_{\beta,\mu} \, ,
\label{A6bis}
\end{equation} 
where the primed expectation value $\langle \, \ldots \,  \rangle'_{\beta,\mu}$ 
indicates that the elements $a^*_{l_1}(k_1)$ and $a_{m_k}(p_k)$ are removed from the string. 
Iterating this step in $\langle \, \cdots \,  \rangle'_{\beta,\mu}$ one obtains 
the Laplace expansion of the determinant in \eqref{A6} along the first row, which concludes the argument. 

{}Finally, one extends \cite{M-11} the correlation functions (\ref{A4}-\ref{A6}) to negative 
momenta by using the reflection-transmission constraints 
\begin{equation}
a_i(k) = \sum_{j=1}^N \S_{ij}(k) a_j(-k)\, , \qquad 
a^*_i(k) = \sum_{j=1}^N a^*_j(-k)\S^*_{ji}(k)\, . 
\label{A7}
\end{equation} 
\bigskip 

\section{Some basic properties of the $\T$-matrix}
\medskip 

The Hermitian matrices $\T^i$, defined by (\ref{T}), play a fundamental role. For illustrating their properties 
it is enough to focus on $\T^1$ associated with lead $L_1$. First of all, 
there exists a $N\times N$ unitary matrix 
$U$ diagonalising $\T^1$. After some algebra one finds 
\begin{equation} 
U\, \T^1\, U^* = \T^1_d  \equiv {\rm diag}[0,0,...,-\sqrt{\tau(\omega)},\sqrt{\tau(\omega)}\, ]\, , 
\label{Tmat1}
\end{equation}
where 
\begin{equation}
\tau(\omega) = \sum_{i=2}^N \tau_i (\omega)\, , \qquad  \tau_i (\omega) \equiv |\S_{i1}(\sqrt{2m\omega})|^2 \, , 
\label{Tmat2}
\end{equation}
is the total transmission probability between the lead $L_1$ and the remaining leads $L_i$ with $2\leq i\leq N$. 
We conclude that $N-2$ of the eigenvalues of $\T^1$ vanish, the nontrivial two being 
$\pm \sqrt{\tau(\omega)}$. A complete system of eigenvectors is given by 
\begin{eqnarray} 
v_1 &=& (0,\, -{\overline \S}_{N1},\, 0,\, ...\, ,0,\,{\overline \S}_{21})\, , \nonumber \\
v_2 &=& (0,\, 0,\, -{\overline \S}_{N1},\, ...\, ,0,\, {\overline \S}_{31})\, , \nonumber \\ 
\nonumber \\
&\cdots& \qquad \cdots \qquad \cdots  \qquad \cdots  \qquad  \nonumber \\
v_{(N-2)} &=& (0,\, 0,\, 0,\, ...\, ,-{\overline \S}_{N1},\, {\overline \S}_{(N-1)1})\, , 
\label{Tmat3}
\end{eqnarray} 
with eigenvalue 0 and 
\begin{equation} 
v_{\mp} = (\pm \sqrt{\tau}-\tau,\, {\overline \S}_{11}\S_{21},\, {\overline \S}_{11}\S_{31},\, ...\, , {\overline \S}_{11}\S_{N1})\, , 
\label{Tmat4}
\end{equation} 
with eigenvalues $\mp \sqrt{\tau}$. Orto-normalising the system (\ref{Tmat3},\ref{Tmat4}), one determines the 
diagonalising matrix $U$ in explicit form.

\bigskip 

\section{Generating function in a single energy channel}
\medskip 

We summarise here the key points of the computation of the expectation value \eqref{fe3} which leads 
to the basic result \eqref{gf6}. Without loss of generality we concentrate on the lead $L_1$, the 
extension to other leads being straightforward. The problem consists in computing 
\begin{equation}
\langle\e^{\ri\lambda j_\omega^1}\rangle_{\beta,\mu}
= \langle \e^{\ri\lambda \sum_{i,j=1}^N a_i^*\T^1_{ij}a_j}\rangle_{\beta,\mu}\, .
\label{C1}
\end{equation}
It is convenient for this purpose to change the basis in the algebra (\ref{alg}) according to 
\begin{equation}
b_i = \sum_{j=1}^N U_{ij}\, a_j \, , \qquad b^*_i = \sum_{j=1}^N a_j^*\, U^*_{ji}\, ,
\label{C2}
\end{equation}
where $U$ is the matrix diagonalising $\mathbb{T}^1$ according to (\ref{Tmat1}). In the new basis 
the expectation value (\ref{C1}) reads 
\begin{equation}
\langle\e^{\ri\lambda j_\omega^1}\rangle_{\beta,\mu}
	= \langle\prod_{j=1}^N\e^{\ri\lambda (\mathbb{T}_d^1)_{jj}b_j^*b_j}\rangle_{\beta,\mu}
	= \langle\prod_{j=1}^N\left[1+\left(\e^{\ri\lambda (\mathbb{T}_d^1)_{jj}}-1\right) b_j^*b_j\right]\rangle_{\beta,\mu} \, .
\label{C3}
\end{equation} 
In the case $N=2$ one has $\mathbb{T}^1_d={\rm diag}(-\sqrt{\tau},\sqrt{\tau})$ and one finds 
\begin{equation}
\begin{split}
\langle\e^{\ri\lambda j_\omega^1}\rangle_{\beta,\mu} 
&	= {\rm det} \left [\II + \left (\e^{\ri \lambda \T^1_d} -\II \right ) 
	\begin{pmatrix}
	\langle b^*_1b_1\rangle_{\beta,\mu} & \langle b^*_1b_2\rangle_{\beta,\mu} \\
	\langle b^*_2b_1\rangle_{\beta,\mu} & \langle b^*_2b_2\rangle_{\beta,\mu}
	\end{pmatrix}
	\right ]
\\
&	= 1+ \ri c_1 \sqrt{\tau}\, \sin (\lambda  \sqrt{\tau}) +c_2 \left [\cos (\lambda  \sqrt{\tau})-1\right ]\, .
\label{C4}
\end{split}
\end{equation} 
Combining (\ref{C4}) with (\ref{gf5},\ref{tgf}) one gets the final result \eqref{gf4}.

The extension to a generic $N>2$ is attained by expanding the product in \eqref{C3} 
and expressing the multiple correlators of the operators $\lbrace b_i,b_i^*\rbrace$ 
in terms of two-point functions through the analogue of formula \eqref{A6}. 
One recognises at this point that the series can be re-summed as a determinant 
\begin{equation}
\langle\e^{\ri\lambda j_\omega^1}\rangle_{\beta,\mu}
	= {\rm det} \left [\II + 
\left (\e^{\ri \lambda \T^1} -\II \right ) \D \right ]\, ,
\label{C5}
\end{equation}
where $\D$ is the diagonal matrix (\ref{m1}) of the Fermi distributions of the reservoirs. 
In the basis in which $\T^1$ is diagonal the result (\ref{C5}) takes the form 
\begin{equation}
\langle\e^{\ri\lambda j_\omega^1}\rangle_{\beta,\mu}
	= {\rm det} \left [\II + 
\left (\e^{\ri \lambda \T_d^1} -\II \right ) U\, \D\, U^*\right ]\, . 
\label{C6}
\end{equation}
Using the explicit form of $U$, which can be deduced from (\ref{Tmat3}, \ref{Tmat4}), 
one obtains from (\ref{C6}) the cumulant generating function (\ref{gf6} - \ref{gf8}) in the multi terminal case.

\end{document}